\documentclass{epl}
\usepackage{amssymb}

\title{Current-phase relation of the SNS junction in a superconducting loop}
\author{Mun Dae Kim\inst{1}\thanks{E-mail: \email{mdkim@kias.re.kr}} \and Jongbae Hong\inst{2}}
\institute{
  \inst{1} Korea Institute for Advanced Study, Seoul 130-722, Korea\\
  \inst{2} School of Physics, Seoul National University, Seoul 151-742, Korea
}


\begin{document}

\maketitle

\begin{abstract}
We study the current-phase relation of the superconductor/normal/superconductor (SNS)
junction imbedded in a superconducting loop. Considering the current conservation  and
free energy minimum conditions, we obtain the persistent currents of the SNS
loop.  At finite temperature we can explain the experimentally observed highly
non-sinusoidal currents which have maxima near the zero external flux.
\end{abstract}

\section{Introduction}

Physically interesting superconducting loops with junctions are the
superconducting quantum interference device (SQUID) and the
superconductor/normal/superconductor loop with long normal sector (${\rm SN_{long}S}$ loop).
Both are mesoscopic size and
yield persistent currents when a magnetic flux threads the loop.
%
The thickness of the Josephson junctions in SQUID loop is much smaller
than the superconducting  coherence length, but the thickness of
normal segment $d$ of ${\rm SN_{long}S}$ loop can be larger than that.
The persistent current in the former system flows by tunneling
the Josephson junction, while that in the latter by the long range
proximity effect.
Since these two loops are very different in nature, we can anticipate
a different current-phase relation for ${\rm SN_{long}S}$ loop, but it was not explicitly demonstrated in
previous studies for the superconductor/normal/superconductor hybrid junction
\cite{Gennes,Kulik,Ivanov} and for the ${\rm SN_{long}S}$ loop \cite{Buttiker,Mtb}
in connection with the Andreev reflection process \cite{Andreev}.


The current described by the motion of a pair
of electron and hole in the normal sector changes into that by a
Cooper pair in the superconducting sector. There is an
intermediate region near the edge of superconductor, where the
current is described by quasiparticles and  quasiholes.
Then the current carried by the electrons and the
holes in the normal sector should be the same as that carried by
quasiparticles and  quasiholes in the intermediate region.
This current conservation condition can be satisfied by
considering that two wave vectors of the quasiparticle and  the quasihole  can be different
from each other like those of electron and hole in the normal sector.
In solving the  Bogoliubov-de Gennes (BdG) equation,
we consider the average energy of  a pair of particles
is a dynamic variable rather than a constant chemical potential.
Thus the values of the dynamic variables are determined
by minimizing the free energy of the ${\rm SN_{long}S}$ loop.

Recently an experiment for the high-$T_c$ supercondutor (HTS) junction with interlayer much thicker
than the superconducting correlation length has been reported \cite{IlichevNS},
where the HTS junction was incorporated in a superconducting loop with threading external magnetic flux $\Phi_{\rm ext}$.
In this single-junction interferometer experiment, the current-phase relation shows the highly
non-sinusoidal behaviors so that the slope of the current for zero external flux  becomes larger
than that for $\Phi_{\rm ext}=\Phi_0/2$ with the superconducting unit flux quantum
$\Phi_0\equiv h/2e$ as temperature goes down.
%
The tunneling currents are supposed to take place through the long range proximity effect across the thick
$\rm{PrBa_2Cu_3O_7}$ (PBCO) interlayer. Since the $\rm{CuO}$ chains in the PBCO layer are metallic
at low temperature, the above junction  can be considered as a ${\rm SN_{long}S}$ junction incorporated in a superconduction loop.
In this study we calculate the persistent current of the above ${\rm SN_{long}S}$ loop considering
the  current conservation and  free energy minimum conditions and
show that the experimentally observed non-sinusoidal type current can emerge through the long-range
proximity effect in the ${\rm SN_{long}S}$ loop.

%
%

\section{Andreev reflections in a superconductor/normal/superconductor loop}

The quasiparticles in an SNS loop with
threading external magnetic flux can be described by the Bogoliubov-de Gennes (BdG) equation
\begin{eqnarray}
\label{BdG}
\left(\matrix{H_0-\xi & \Delta (z) \cr
\Delta^* (z) & -H^*_0+\xi}\right)
\left(\matrix{u(z) \cr v(z)}\right)
=E\left( \matrix{u(z) \cr v(z)}\right),
\end{eqnarray}
where $H_0=(-i\hbar\partial/\partial z-eA/c)^2/2m_e$ with the electron mass  $m_e$,
$A=\Phi_{\rm ext}/L$ is the vector potential with the circumference of loop $L$,
and $\Delta(z)$ is the pair potential as a function of spatial coordinate $z$.

For a superconducting loop {\it without} junction threaded by an
Aharonov-Bohm flux, the quasiparticle wave functions $u(z)$ and
$v(z)$  contain extra factor due to gauge invariance such as
$u(z)=C e^{i(\lambda_{e}+\pi f/L)z}$ and $v(z)=C e^{i(\lambda_{h}-\pi f/L)z}$,
where $f\equiv \Phi_{\rm ext}/\Phi_0$.
Uniform flow of persistent current is derived by the BdG equation with the pair potential,
$\Delta (z)=\Delta e^{i(\lambda_{e}-\lambda_{h}+\pi f/L)z}$ \cite{Gennes}.

\begin{figure}[t]
\vspace*{-0.3cm}
\onefigure[scale=0.5]{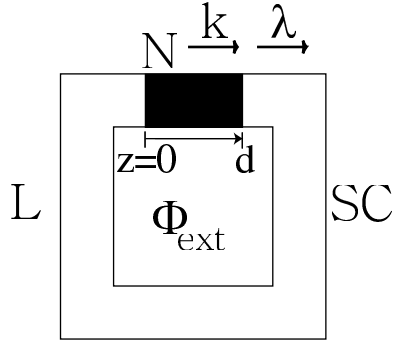}
\vspace*{0.3cm}
\caption{A superconductor/normal/superconductor loop with length $L$. $k$ and $\lambda$
denote the wave vector of a electron in normal sector and a quasiparticle
in superconducting sector, respectively.}
\label{fig:loop}
\end{figure}

If the superconducting loop is interrupted by a normal sector as shown in Fig. \ref{fig:loop},
the wave function $\Psi=(u(z)~~ v(z))^T$ of a pair of electron and hole in normal sector $(0<z<d)$
and of a pair of quasiparticle and quasihole in the intermediate region of superconducting sector
$(d\lesssim z \, \, \mbox{and} \, \, z\lesssim L)$ is given by
\begin{eqnarray}
\label{wfns}
\left( \matrix{u(z) \cr v(z)}\right)=\left\{
\matrix{
\left( \matrix{A e^{i(k_0+\frac{\pi}L f)z} \cr B e^{i(k_1-\frac{\pi}L f)z}
}\right)~~~~~(0<z<d)~ \cr
\left( \matrix{C e^{i(\lambda_{+e}+\frac{\pi}L f)z+i\eta} \cr
C  e^{i(\lambda_{+h}-\frac{\pi}L f)z-i\eta}}\right)~~~~~ (d \lesssim z)~ \cr
\left( \matrix{D e^{i(\lambda_{-e}+\frac{\pi}L f)z-i\eta} \cr
D e^{i(\lambda_{-h}-\frac{\pi}L f)z+i\eta}}\right)~~~~~ (z \lesssim L),
} \right.
\end{eqnarray}
where $\eta$ is the phase shift due to Andreev reflection at the NS interface
\cite{Buttiker,Mtb,KulikComm} and, when a pair of quasiparticles passes through the NS interface,
each quasiparticle acquires the additional phase equal to $\eta$.
Since we do not make any assumption about the sign of the wave vectors $k_0$ and $k_1$,
this wave function can describe  both the excitations moving clockwise and counterclockwise.
Here  a notable point is that the wave vectors  of the quasiparticle and the quasihole
in the previous work \cite{Kulik,Buttiker,Mtb} are set to be the same.
It is, however, natural to discriminate the wave vectors like those of particles
$k_0$ and $k_1$ in the normal sector in order to satisfy the current conservation
condition. Therefore, we introduce different wave vectors, $\lambda_e$
and $\lambda_h$, for the quasiparticle and the quasihole in the intermediate region of
superconducting sector.

This equation can be solved easily for the normal sector in which $\Delta(z)=0$,
and yields the relations for $\xi$ and $E$ such that
\begin{eqnarray}
\label{normalU}
\xi=\frac{\hbar^2}{4m_e}(k^2_0+k^2_1),\\
\label{normalE}
E=\frac{\hbar^2}{4m_e}(k^2_0-k^2_1).
\end{eqnarray}
%
In the intermediate region of superconducting sector, however,
the BdG equation must be solved with the pair potential
\begin{eqnarray}
\Delta(z)=\left\{\matrix{
\Delta e^{i(\lambda_{+e}-\lambda_{+h}+2\pi f/L)z}~(z\gtrsim d)~ \cr
\Delta e^{i(\lambda_{-e}-\lambda_{-h}+2\pi f/L)z}~(z\lesssim L).
}\right.
\end{eqnarray}
Then the BdG equation for $z\gtrsim d$ becomes
$\hbar^2\lambda^2_{+e}/2m_e-\xi+\Delta e^{-2i\eta}=E$ and
$\Delta e^{2i\eta}-\hbar^2\lambda^2_{+h}/2m_e+\xi=E.$
Representing $\lambda_{+e}$ and $\lambda_{+h}$
such as $\lambda_{+e}=\lambda_0+i\lambda'_0$ and
$\lambda_{+h}=\lambda_1+i\lambda'_1$ we get an expression for $\xi$ and $E$,
\begin{eqnarray}
\label{U}
\xi&=&\frac{\hbar^2}{4m_e} (\lambda^2_0+\lambda^2_1)(1-\alpha)\\
\label{E}
E&=&\frac{\hbar^2}{4m_e}
(\lambda^2_0-\lambda^2_1)(1+\alpha)+\Delta\cos2\eta
\end{eqnarray}
with $\alpha \equiv(m_e\Delta \sin 2\eta/\hbar^2\lambda_0\lambda_1)^2$,
$\lambda'_0=m_e\Delta \sin 2\eta/\hbar^2\lambda_0$ and
$\lambda'_1=m_e\Delta \sin 2\eta/\hbar^2\lambda_1$.
For $z\lesssim L$, we can also solve the BdG equation and find that
$\lambda_{-e}=\lambda_0-i\lambda'_0$ and
$\lambda_{-h}=\lambda_1-i\lambda'_1$.

The phase matching conditions for the wave function
of Eq. (\ref{wfns}) at $z=d$ and $z=L$ are given by
${\hat A} e^{ik_0 d-i\lambda_0 d-i\eta}={\hat B} e^{ik_1 d-i\lambda_1 d+i\eta}={\hat C}$ and
${\hat A} e^{-i(\lambda_0+\pi f/L)L+i\eta}={\hat B} e^{-i(\lambda_1-\pi f/L)L-i\eta}= {\hat D}$,
where ${\hat A}, {\hat B}, {\hat C}$ and ${\hat D}$ represent the
phase part of the coefficients, $A,B,C$ and $D$, respectively. The
condition for the existence of a solution leads to the boundary
condition
\begin{eqnarray}
\label{contns}
\left(k_0-k_1+\frac{2\pi f}L \right)d+\left(\lambda_0-\lambda_1+\frac{2\pi f}L \right)(L-d)
-4\eta =2\pi n.
\end{eqnarray}
%
When an electron becomes Andreev-reflected at the superconductor/nomal interface,
the transmitted quasiparticle pair obtains the additional phase $2\eta$.
The phase $4\eta$ in the boundary condition in Eq. (\ref{contns}) is the sum of
two phase changes $2\eta$ due to the Andreev reflections at each interface.

The condition of current conservation at NS interface will be given
by using the representation of the flux
$j=-(1/2m)[\phi(-i\hbar\partial/\partial z-eA/c)\phi^*
-\phi^*(-i\hbar\partial/\partial z-eA/c)\phi ]$.
In the wave function $\Psi=(u(z) ~~ v(z))^T$ in Eq. (\ref{wfns}), $u(z)$
is the electronlike wave function and $v(z)$ is the holelike wave
function which is the complex conjugate of the electronlike wave function.
The current representation for a pair of electrons thus should be obtained
with the wave function $\phi=(u(z) ~~ v^*(z))^T$ and we can get the relation
\begin{eqnarray}
\label{current}
k_0-k_1=\lambda_0-\lambda_1 \equiv 2q.
\end{eqnarray}
%
The Cooper pairs in superconducting sector then should
carry the current $I\equiv n_c 2e\hbar 2q/m_c$
with the density of Cooper pairs $n_c$ and $m_c=2m_e$.

\section{Persistent currents of a superconductor/normal/superconductor loop}

When we  calculate the current through the ${\rm SN_{long}S}$ junction,
we may be able to set the average energy $\xi$ of the particle and the hole in Eq. (\ref{normalU})
equal to the chemical potential. But, in the ${\rm SN_{long}S}$ loop,
$\xi$ in Eq. (\ref{normalU}) need not be the constant chemical potential.
For example, we consider a simple normal loop with a threading flux.
The average energy of two particles $\xi$ at Fermi level is different from the chemical potential $\mu$
such that  $\xi=(1/2)(\hbar^2/2m_e) (k_0^2 +k_1^2)=(1/2)(\hbar^2/2m_e) ((k_F-\pi f/L)^2 +(k_F+\pi f/L)^2)
\neq (\hbar^2/2m_e)k^2_F=\mu$.
Since $k_0$ and $k_1$ depend on the dynamic variable $\eta$ as well as  $f$ in the ${\rm SN_{long}S}$ loop,
$\xi$ cannot be set as a constant chemical potential but should also be a dynamic variable to be determined.

Since an extra variable $\xi$ is introduced, we need one more independent relation.
That is given by the requirement of free energy minimum \cite{Kim}.
Since the intermediate region of superconducting sector is so thin that
we neglect the energy of this region in the free energy expression and consider only the
energy of Cooper pairs.
In superconducting sector the Cooper pairs carry the persistent current corresponding
to the Cooper pair wave vector $2q=\lambda_0-\lambda_1$ and the energy of a Cooper pair
can be written as $(\hbar^2/2m_c)(2q)^2$ with $m_c=2m_e$.
Therefore the total free energy per particle $U_{\rm tot}$ can be written as
\begin{eqnarray}
2U_{\rm tot}=\frac{\hbar^2}{2m_e} (k^2_0+k^2_1)\frac{d}{L}
+\frac{\hbar^2}{2m_c}(\lambda_0-\lambda_1)^2 \left(1-\frac{d}{L}\right).
\end{eqnarray}
Using Eqs. (\ref{normalU}) and (\ref{U}), the total free energy can be represented in $\lambda_0$
and $\lambda_1$.

From the condition $dU_{\rm tot}/d\eta=0$, we obtained the equation,
\begin{eqnarray}
\label{CE}
&&\left[(\lambda_0+\lambda_1)+(\lambda_0-\lambda_1)\frac{L}d+\frac{(2m_e\Delta\sin
2\eta)^2}{2\lambda^3_0 \hbar^4}\right]\frac{d\lambda_0}{d\eta}\nonumber\\
&+&\left[(\lambda_0+\lambda_1)-(\lambda_0-\lambda_1)\frac{L}d+\frac{(2m_e\Delta\sin
2\eta)^2}{2\lambda^3_1 \hbar^4}\right]\frac{d\lambda_1}{d\eta}\nonumber\\
&=&\frac{\lambda^2_0+\lambda^2_1}{2\lambda^2_0\lambda^2_1}
\left(\frac{2m_e\Delta}{\hbar^2}\right)^2\sin 4\eta.
\end{eqnarray}
Here $d\lambda_0/d\eta$ and $d\lambda_1/d\eta$ can be obtained by differentiating
Eqs. (\ref{normalU}), (\ref{normalE}), (\ref{U}), (\ref{E}), (\ref{contns}) and (\ref{current}) such as
$d\lambda_0/d\eta = [{\mathcal F}+\{8\lambda_1(\alpha k_0+k_1)+4\lambda_0\Gamma\sin 2\eta\}/L]/{\mathcal D}$ and
$d\lambda_1/d\eta = [{\mathcal F}+\{8\lambda_0(\alpha k_0+k_1)-4\lambda_1\Gamma\sin 2\eta\}/L]/{\mathcal D}$,
where ${\mathcal F}\equiv 2\lambda_0\lambda_1\Gamma\cos 2\eta
-2\tilde{\Delta}(k_0+k_1)\sin 2\eta-8k_0k_1/L$
and  ${\mathcal D} \equiv -2(\lambda_0-\lambda_1)(\alpha k_0+k_1)
+(\lambda_0+\lambda_1)\Gamma \sin 2\eta$
with $\Gamma\equiv (k_0\lambda^2_0-k_1\lambda^2_1)\tilde{\Delta}^2\sin 2\eta
/2\lambda^3_0\lambda^3_1$ and $\tilde{\Delta}\equiv 2m_e\Delta/\hbar^2$.

\begin{figure}[t]
\vspace*{-6.5cm}
\onefigure[scale=0.9]{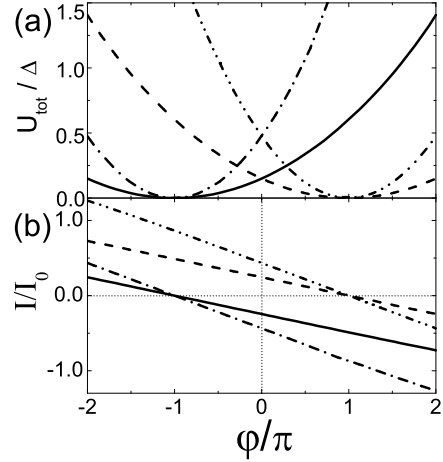}
\caption{(a) Excitation spectrum of the ${\rm SN_{long}S}$ loop. The solid (dashed)
line for $\varphi<0$ ($\varphi>0$) shows the ground state energy, where $\varphi\equiv 2\pi f$.
(b) Persistent currents corresponding to the energy levels in (a).
Here we set $L\sqrt{\tilde{\Delta}_0}=10$  and $\Delta/\Delta_0=1.$}
\label{fig:EnDiagram}
\end{figure}

Solving the coupled equations (\ref{normalU}), (\ref{normalE}), (\ref{U}), (\ref{E}), (\ref{contns}),
(\ref{current}) and (\ref{CE}) numerically, we obtain the energy levels corresponding to the  two solutions
of the BdG equation in Eq. (\ref{BdG})
as a function of external flux $\varphi\equiv 2\pi f=2\pi\Phi_{\rm ext}/\Phi_0$
as shown in Fig. \ref{fig:EnDiagram}(a), where the solid and the
dashed lines denote the lower levels and the other lines the
higher levels. The ground state corresponds to the solid (dashed)
line for $\varphi <0 ~(\varphi >0)$. The persistent currents in Fig.
\ref{fig:EnDiagram}(b) are represented by the same lines as those
of the corresponding states in Fig. \ref{fig:EnDiagram}(a). The
persistent current, $I\equiv n_c 2e\hbar 2q/m_c$, can be written
by $I=I_0 2q/\sqrt{\tilde{\Delta}_0}$, where  $I_0\equiv n_c
2e\hbar\sqrt{\tilde{\Delta}_0}/m_c$ and $\tilde{\Delta}_0\equiv
2m_e\Delta_0/\hbar^2$ is a gap potential chosen arbitrary.

Since the Cooper pairs in the superconductor are in the coherent condensate state,
each Cooper pair carries the same superconducting current.
A Cooper pair changes into a pair of normal electrons in the normal sector via
a pair of quasiparticles in the intermediate region.
Since the current in the loop should be conserved, the current equal to the macroscopic persistent
current in the superconducting sector flows in the normal sector.
%
In Fig. \ref{fig:EnDiagram}(b), we can see  several  persistent currents,
of which the persistent current at ground state corresponds to the solid line for $\varphi<0$
and the dashed line for $\varphi>0$, the saw-tooth type current.

Recently a single-junction interferometer experiment was done on HTS junction \cite{IlichevNS}
(${\rm YBa_2Cu_3O_{7-x}}$ /${\rm PrBa_2Cu_3O_7}$/${\rm YBa_2Cu_3O_{7-x}}$)
which is incorporated into a superconducting loop with penetrating magnetic flux.
Since  there is no  misorientation angle between the two d-wave superconductors
across the interlayer in this experiment, the phase difference across the interlayer
can be brought about only by the threading magnetic flux.
They obtained the current-phase relation  at finite temperature,
where the slope for zero external flux is larger  than that for the external flux $\varphi=\pi$.

\begin{figure}[t]
\vspace*{-1cm}
\onefigure[scale=0.8]{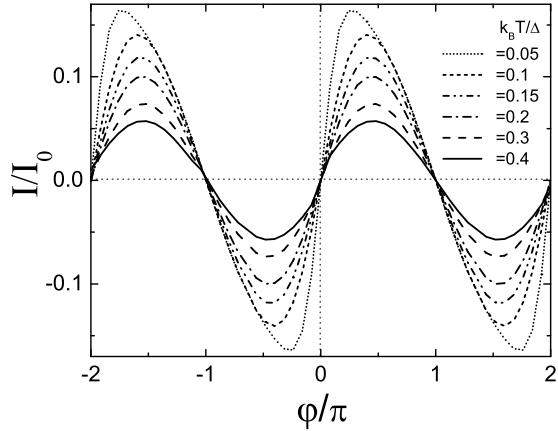}
\caption{Current-phase relation at finite temperatures. As the temperature goes down,
the maximum approaches the point $\varphi=0$}
\label{fig:PcSin}
\end{figure}

In this experiment, the thickness of the PBCO interlayer of the junction is as long as hundreds ${\rm \AA}$
which is much larger than the high-$T_c$ superconducting correlation length by
an order of magnitude and thus the Cooper pairs cannot directly tunnel the junction.
These long range proximity effects have been observed in many experiments on HTS junctions
\cite{Barner,Bozovic,Polturak,Umezawa} and considered as a characteristic of ${\rm SN_{long}S}$ junctions.
Thus the experimental results may be explained by calculating the currents
of the ${\rm SN_{long}S}$ loop as a function of  the threading external flux.

Here we can raise a question that whether it is a real proximity effect, i.e., there are
filaments of pinholes or microshots so that the Cooper pairs can be transported by the resonant
tunneling through localized states in filaments.
Recently an experiment on the trilayer HTS junction \cite{Bozovic} has been reported,
where they synthesized atomically smooth films of HTS and uniform trilayer junction
so that they conclude that the long-range proximity effect do not originate from the resonant
tunneling through the energy-aligned states in the microshots, but it is an intrinsic
property of the interlayer.
In fact, the supercurrents are known to flow via metallic ${\rm CuO}$ chains in the PBCO interlayer
\cite{Fehrenbacher,Suzuki}.



From the currents in Fig. \ref{fig:EnDiagram}(b) corresponding to the states in Fig. \ref{fig:EnDiagram}(a)
we can obtain the persistent currents of the ${\rm SN_{long}S}$ loop at finite temperatures.
At finite temperature, since the current state with energy $\epsilon$ has the probability proportional to
$e^{-\epsilon/k_B T}$, we can obtain the thermally averaged persistent currents
from those current states in Fig. \ref{fig:EnDiagram} as shown in Fig. \ref{fig:PcSin}.
We can observe highly non-sinusoidal currents in Fig.  \ref{fig:PcSin},
where the amplitude maxima approach the point $\varphi=0$ and the slope of current
at $\varphi=0$ becomes larger than that at $\varphi=\pi$  as temperature goes down.
The current takes the sinusoidal form only after the temperature goes up such that
$k_B T/\Delta \approx 0.4$ which corresponds to the temperature near $T_c$.

Actually the similar behavior can also be seen in the loop interrupted by a grain boundary Josephson junction
not by an ${\rm SN_{long}S}$ junction, which however comes from the phase difference across
the grain boundary Josephson junction due to the misorientation angle between two d-wave superconductors \cite{IlichevJJ}.
In ${\rm SN_{long}S}$ junctions the wave functions acquire the Andreev reflection phase shift $\eta$
which appears in the boundary condition of Eq. (\ref{contns}).  The Andreev phase shift takes the place of the
misorientation angle of the grain boundary Josephson junction and thus results in the non-sinusoidal current-phase relation as
shown in the manuscript.
In Ref. \cite{IlichevNS} the authors explained their own experimental
results by assuming thermal fluctuations which induced the highly
non-sinusoidal current-phase relation at low temperatures. At
higher temperatures near the critical temperature they can recover
the sinusoidal current-phase relations. In the present study,
however, we can show that the non-sinusoidal current-phase
relation emerges naturally at low temperature without assuming
thermal fluctuations.

\section{Summary}

We have studied  the current-phase relation of  the ${\rm SN_{long}S}$ loop with threading magnetic flux.
The Cooper pairs in the superconducting sector become the electron pairs in the normal sector
via the quasiparticle states in the intermediate region.
The net current in the ${\rm SN_{long}S}$ loop should be preserved and thus we introduce  different wave vectors
of the pair of quasiparticles in order to satisfy the current conservation condition.
Furthermore, since the average energy of a pair of particles should be a dynamic variable,
we also consider the free energy minimum condition.
We obtained the persistent currents of the ground and excited states and
found that we can explain the experimentally observed highly non-sinusoidal
current-phase relation with maxima near the zero external flux at low temperature.





\acknowledgments
This work was supported by Korea Research Foundation Grant
No. KRF-2003-070-C00020.


\begin{thebibliography}{0}
\bibitem{KulikComm} In Ref. [2], this phase shift can be defined as $\eta\equiv\phi/2$ with $e^{-i\phi}=\gamma$.
\end{thebibliography}
\end{document}